\documentclass[10pt,journal,compsoc]{IEEEtran}

\usepackage[nocompress]{cite}

\usepackage[utf8]{inputenc} 
\usepackage{amsmath, amsthm, amssymb}
\usepackage{graphicx} 
\usepackage{subfigure}

\DeclareGraphicsExtensions{.png}
\DeclareMathOperator{\sgn}{sgn}
\newcommand{\idf}{\mathrm{idf}}
\newcommand{\vecTau}{\overrightarrow{\boldsymbol{\tau}}}
\newcommand{\vecBeta}{\overrightarrow{\boldsymbol{\beta}}}
\newcommand{\hbAvrg}[1]{\bar{#1}}
\newcommand{\hbVect}[1]{\mathbf{#1}}
\newcommand{\hbMtrx}[1]{\mathbf{#1}}
\newcommand{\hbFigSize}{0.9}

\newcommand{\reffig}[1]{Fig.~\ref{#1}}

\newcommand{\reftbl}[1]{Table~\ref{#1}}
\newcommand{\refsec}[1]{Sec.~\ref{#1}}
\newcommand{\refcite}[1]{Ref~\cite{#1}}
\newcommand{\hbEtal}{et al. }
\newcommand{\hbIe}{i.e., }
\usepackage{xcolor}
\usepackage[iso]{datetime}
\usepackage[colorlinks=true,linkcolor=red,urlcolor=blue,citecolor=red]%
	{hyperref}

\begin{document}

\title{
	Context Sensitive Article Ranking with Citation Context Analysis
}

\author{Metin Doslu
        and~Haluk O. Bingol%
	\IEEEcompsocitemizethanks{%
		\IEEEcompsocthanksitem %
		M. Doslu and H.O. Bingol are with 
		the Department of Computer Engineering,
		Bogazici University, Istanbul\protect\\
		E-mail: metindoslu@gmail.com and bingol@boun.edu.tr
	}%
}%

\IEEEtitleabstractindextext{%
\begin{abstract}
	It is hard to detect important articles in a specific context. 
	Information retrieval techniques based on full text search can be 
	inaccurate to identify main topics and 
	they are not able to provide an indication about 
	the importance of the article. 
	Generating a citation network is a good way 
	to find most popular articles but this approach is not context aware. 
	
	The text around a citation mark is generally 
	a good summary of the referred article. 
	So citation context analysis presents an opportunity 
	to use the wisdom of crowd for 
	detecting important articles in a context sensitive way. 
	In this work, we analyze citation contexts 
	to rank articles properly for a given topic. 
	The model proposed uses citation contexts in order to create 
	a directed and edge-labeled citation network based on the target topic. 
	Then we apply common ranking algorithms in order 
	to find important articles in this newly created network.
	We showed that this method successfully detects 
	a good subset of most prominent articles in a given topic. 
	The biggest contribution of this approach is that 
	we are able to identify important articles for a given 
	search term even though these articles do not contain this search term.
	This technique can be used in other linked documents including 
	web pages, 
	legal documents, and
	patents as well as scientific papers.
\end{abstract}

}

\maketitle

\IEEEdisplaynontitleabstractindextext
\IEEEpeerreviewmaketitle

\section{Introduction}

A researcher needs to know about related work 
before starting to study on a topic. 
In this context, 
citation indexes such as CiteSeerX~\cite{%
	urlCiteSeetX}
are very useful to navigate through related research articles.
Some of the citation indexes provide 
a medium to search over full text of articles.
Citation indexes are also able to index articles 
without access to full text with the help of articles cite them. 
They also provide a way to evaluate importance of an article 
because they also report the number of times the article is cited.

\begin{figure}[htbp] 
\begin{center}
	\fbox{
		\includegraphics[width=.95\columnwidth]
			{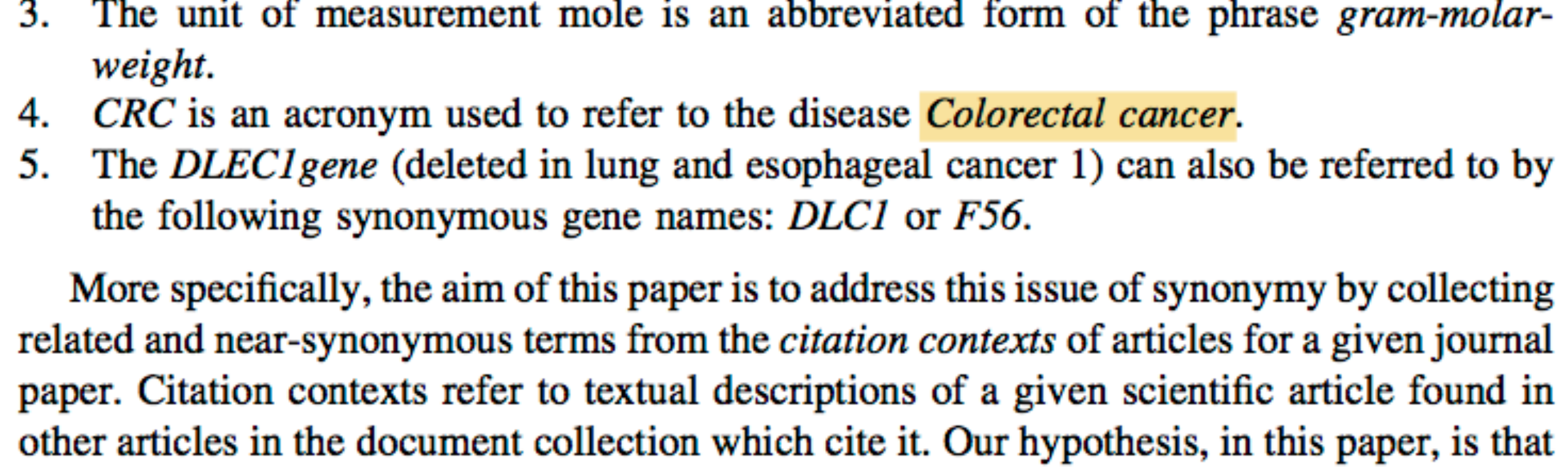}
	}
	\caption{Part of an article about information retrieval.}
	\label{fig:information_retrieval_drawback}
\end{center}
\end{figure}

However it is still an exhaustive work to scan scientific literature 
to find important articles in the interested topic.
Text of an article would contain lots of words 
not related with its main topic. 
These words would be used in examples, controversy arguments etc. 
Search methods 
which use indexing techniques on full text 
suffer from these problems. 
\reffig{fig:information_retrieval_drawback} 
shows a part from an 
article~\cite{%
	Aljaber2009} 
which is about information retrieval. 
This text part contains a term about ``cancer''. 
So any full text indexing technique will index this article for the term ``cancer'' although this term is not related with the main argument of the article.
Indexing techniques also do not have any information about the importance of the target articles.

Citations provide a way to measure the relative impact of articles.
A \emph{citation network} is formed by articles as nodes, 
and there is an arc from article $i$ to article $j$ 
if and only if 
$i$ contains a citation to $j$. 
Forming a citation network and ranking articles according to 
their in-degree, i.e. \emph{citation count}, on this network 
helps to identify important articles over all articles. 

Citation networks provide a way to detect 
the most outstanding articles in the scientific literature. 
However ranking obtained from a citation network 
does not contain context information, 
so one still needs to differentiate them according to the context. 
Combining indexing techniques with a citation network 
would be a good approach to try, 
but this hybrid method still suffers from terms inside the article 
but not related with the main topic of the article. 
Based on results of academic search engines, 
we would infer that they use such hybrid systems 
which consist of full text indexing and 
citation network as a part of their systems.

A \textit{citation context} is essentially 
the text surrounding the reference markers 
used to refer to other scientific 
works~\cite{%
	Aljaber2009}. 
Citation context provides an useful way 
to identify the main contributions of an scientific publication 
because authors refer articles by briefly presenting 
main points of the cited article in citation context. 
To be cited in an article with specific terms is 
a significant indication of importance in that topic. 
More the article is cited with the same terms 
means that this article is more important 
in the topic which these terms represent. 

Citation context is generally formed by the explicit and definitive words that the citing author uses to describe the cited work. 
Most of the time, the citation context is a very good summary of the cited article or points to some important highlights about it. 
In other words, citation context contains representative keywords for the cited work. 
Citation context analysis provides us 
an opportunity to reason about main topics of the cited article 
even though we do not have the contents of the article. 
Schwartz and Hearst stated that it is hypothesized that through time, 
the citation context can more accurately describe 
the most important contributions of an article than its original 
abstract~\cite{%
	Schwartz2006}. 

There are many documents types that contain links.
Patents or
laws refer to some other patents or laws.
It is observed that the link structure formed by citations 
is analogous to that of the web 
where the links are hyperlinks between web pages. 
This motivated us to use widely used network analysis methods for web link analysis 
such as 
HITS~\cite{%
	Kleinberg1999} 
and 
PageRank~\cite{%
	Brin1998} 
to apply on our dataset.

\subsection{Our Approach}

Our contribution is in applying Complex Networks techniques to 
network of linked documents.
For a given query,
a new set of documents retrieved,
a new network is generated on the retrieved documents, 
and standard network based ranking algorithms is run 
on the obtained network.
As a result,
a new ranking is obtained.
Therefore, 
our goal is not build an Information Retrieval (IR) system
but
to develop some new approaches, 
which can be used in an IR system.

We assume that terms in the citation context are already extracted.
Using terms,
we construct context aware citation networks.
Then we present a method on these networks 
to find important articles for a given topic
by using common network analysis methods. 

Utilizing citation contexts helps us to find articles especially in the following situations where there is no way to pinpoint them using full text indexing methods:

\begin{itemize}
	
	\item 
	Suppose that an article proposed a concept, 
	and later someone else build another concept on the top of this concept. 
	In such situations, 
	if you are looking for important articles for the second concept, 
	you would also like to see articles related to the first concept.

	For example, 
	``Hadoop'' was derived from Google File System 
	(GFS)~\cite{%
		Ghemawat2003} 
	and Google's 
	``MapReduce''~\cite{%
		Dean2004} 
	articles. 
	In this case, 
	if one is looking for important articles for ``Hadoop'', 
	he should also able to see articles related to ``MapReduce''
	although ``Hadoop'' is never mentioned in it.
	
	\item 
	If there are closely related concepts, say $c_{1}$ and $c_{2}$.
	It is possible that 
	one document talks about $c_{1}$ 
	while never mentions $c_{2}$.
	In full text approach, there is no way to get documents on $c_{2}$.
	
	For example,
	concepts ``power law'' and ``small world'' in Complex Networks 
	are closely related.
	There can be some articles which talk about ``small world''
	without mentioning ``power law''. In such situations, 
	if one looks for important articles for ``power law'', 
	he also would like to see articles related with ``small world''.
\end{itemize}

\section{Related Work}

We are creating a context aware network by using citation context, 
so we utilized existing work on citation context and network generation with  scientific papers. 
We will summarize related work in this section.

\subsection{Citation Context}

Bradshaw used citation contexts to index cited papers 
in his \emph{Reference Directed Indexing (RDI)} 
method~\cite{%
	Bradshaw2003}. 
The main motivation behind RDI is that, 
in a citation context authors describe a cited article 
with similar terms to a search query used to search it. 
He used a citation context length
that is approximately 100 words long with 
50 words on either side of the point of citation. 
Then he created a list of index terms for the cited article 
from the citation contexts where it is cited. 
As the number of citation contexts to the paper increases, 
a pattern emerges.
Some terms are observed more frequently.
The score of these index terms are increased.
After term indexes for all articles are created in the dataset, 
for a given query,
RDI first checks articles 
which contain all the terms of the query in their index list, 
and then ranks them according to their index scores. 
The results are tested by checking how many relevant documents returned by 
his search engine based on RDI in the top ten and 
compared their results with a full text similarity based index search method.

Research of Bradshaw is the closest study to ours. 
We also used citation context to rank papers for specific topics. 
We differ in two main aspects which will be discussed shortly.
(i)~We created a directed network of articles 
where arcs are labelled by terms obtained from the citation contexts 
instead of indexing citation context.
(ii)~Another contribution of ours is that 
instead of just using search term 
we also search for similar terms in citation contexts 
while forming our citation network. 
This increases accuracy and robustness of our system.

Connected document structure is observed not only in scientific papers 
but in web pages, too.
Ritchie~\hbEtal discussed similarities between 
the web and scientific literature, 
making an analogy as hyperlinks between web pages 
alongside citation links between 
articles~\cite{%
	Ritchie2006}. 
They mentioned that there are fundamental differences like 
greater variability of web pages and 
the independent quality control of scientific texts 
through the peer review process. 
They stated that the analogy between hyperlinks and citations is not perfect 
because the number of hyperlinks varies from web page to web page 
where the number of citations in papers is somehow restricted. 
Aljaber~\hbEtal also makes an analogy between 
citation context in scientific articles and 
anchor text in web 
pages~\cite{%
	Aljaber2009}.

It is shown that citation contexts can be used to cluster 
documents~\cite{%
	Aljaber2009}. 
For each article,
a citation term representation is generated 
from all its citation contexts found in the dataset. 
Then the representation is used to cluster articles.

It is also shown that citation contexts can be used to summarize 
articles~\cite{%
	Qazvinian2010}. 
They extracted significant key phrases from the set of citation contexts 
where key phrases are expressed using $n$-grams. 
Then, they used these key phrases to build the summary.

In a very previous work, 
citations of scientific articles are classified according to 
whether they are 
conceptual or operational, 
organic or perfunctory, 
evolutionary or juxtapositional, 
confirmatory or negational~\cite{%
	Moravcsik1975}.

Although there are variety of works that focus on citation contexts, 
these efforts were relatively on small datasets. 
For example Bradshaw used 10,000 
articles~\cite{%
	Bradshaw2003} 
and 
Ritchie~\hbEtal used 9,000 
articles~\cite{%
	Ritchie2006}. 
Most of the bigger datasets are not well structured and require lots of preprocessing and manual work. 
Problems coined are generally unsupervised and 
evaluation of results requires manual work. 
This makes infeasible to evaluate large result sets. 
For example, in order to evaluate results Bradshaw listed top 10 articles for every test query 
they run on both their system and comparison systems. 
Then, they mixed results and manually checked the relevance of articles 
in the result sets without knowing 
which system found which 
articles~\cite{%
	Bradshaw2003}.

According to Aljaber~\hbEtal 
using terms around citation references with a predefined window size 
is a simple but effective way to determine useful 
terms~\cite{%
	Aljaber2009}. 
They tried different window sizes and found 
that 50 words before and after the citation reference is optimal citation context size for document clustering on their datasets. 
Similarly Bradshaw also used citation contexts of 
100 words length extracted from articles with 
50 words on both sides of the citation mark
to index the cited 
articles~\cite{%
	Bradshaw2003}.

\subsection{Network Generation}

One can use different methods to create a network over an article dataset.  
 
The main assumption behind bibliographic coupling, introduced by Kessler, 
is that similar articles have similar 
references~\cite{%
	Kessler1963}. 
Two articles are 
\emph{bibliographically coupled} 
if and only if 
they cite the same article. 
Number of common citations can be used to create an undirected weighted edge between these two articles. 
 
Another way to create a network from articles is to use co-citation analysis. 
The \emph{co-citation count} for two articles $A$ and $B$ is 
the number of articles 
that cite $A$ and $B$ 
together~\cite{%
	Small1973}. 
We can generate a weighted undirected network 
by creating edges between articles using co-citation counts. 
The main assumption behind co-citation analysis is that similar articles are cited together more frequently. 
Gipp~\hbEtal introduced an extended approach, 
called \emph{Co-citation Proximity Analysis (CPA)}, 
on the top of co-citation 
analysis~\cite{%
	Gipp2009}. 
CPA considers the proximity of citations within an article with basic assumption 
that two articles are more similar 
if citations to them appear closely. 
Then, we can calculate weight of an edge between two articles with 
a function of proximity of citations.

\section{Methodology}
\label{sec:Methodology}

We need a methodology which is good both for relevance and significance. 
An article identifies the main contributions of the cited article and uses related terms when citing this article. 
This gives us invaluable information about the relevance of cited articles with the interested topic. 
Heavily cited articles with related terms generally mean important contributions in the topic of interest, 
so more the citation count means more significant the cited article.

The citation context of a citing article may have many possible meanings: 
it may be off topic or it may convey criticism rather than approval. 
It is hard to determine the intent of the citation context 
automatically~\cite{%
	Abujbara2013}. 
But in aggregate, if an article is cited by many articles with the same terms, 
then it is receiving a kind of collective confirmation 
in the area of the term represents. 
We can extract cumulative understanding of the crowd for 
the cited article from cumulative citation contexts of citing articles.

Our method is not on term extraction.
So we assume that 
we have some means to extract terms from citation context.
Term extraction approach used in our proof of concept implementation 
is discussed later in  \refsec{sec:TermIdentification}.

We propose a new system 
that takes a term as input,
where term can be a single word or multiple words. 
As any other system,
the system
(i)~returns an unordered list of documents 
and
(ii)~orders the list.
The details are given shortly.

\subsection{Citation Network}

\begin{figure}
\begin{center}
	\includegraphics[width=.95\columnwidth]
		{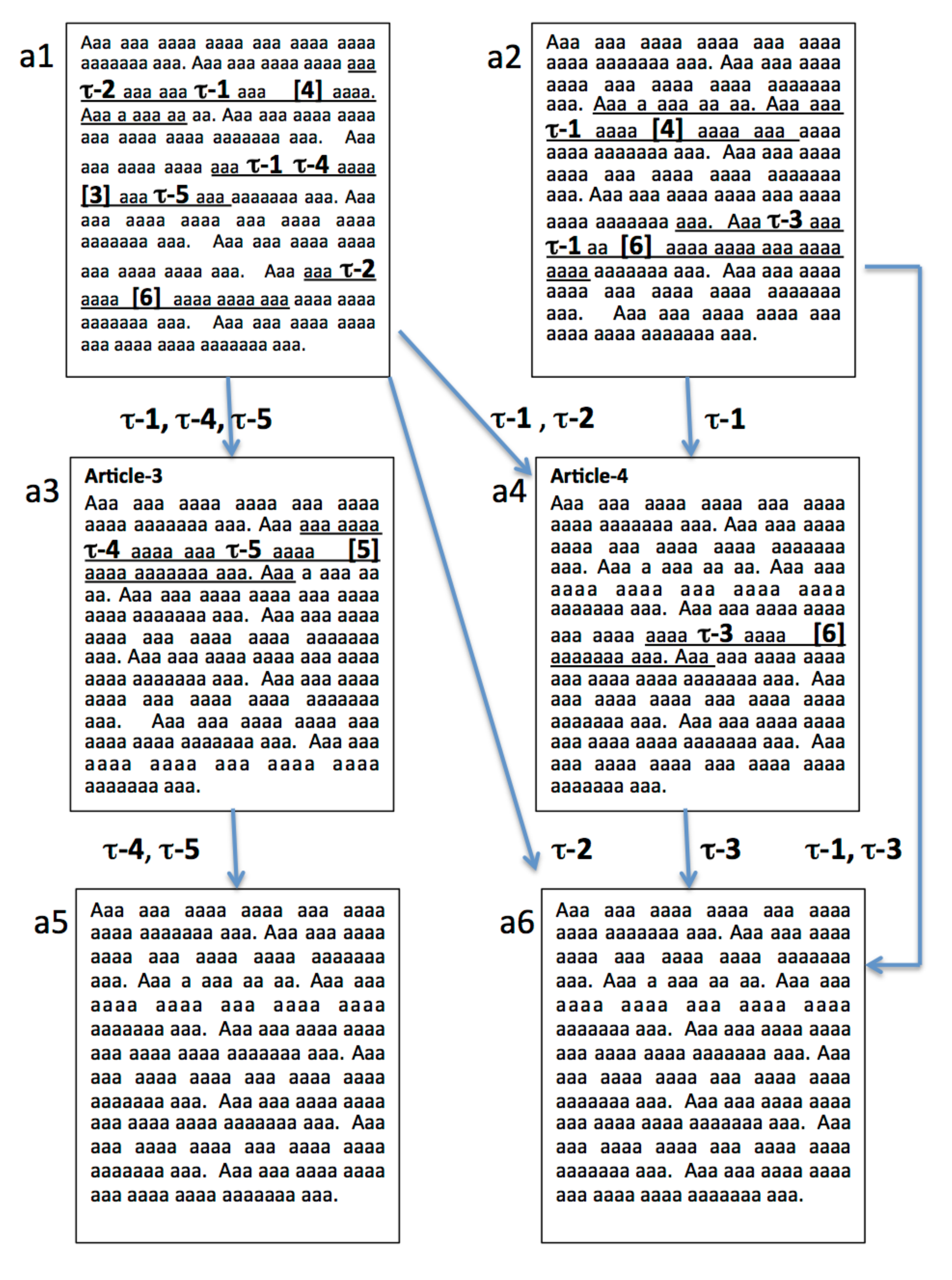}
	\caption{
		Citation network of a set of articles. 	
		For convenience, 
		the citation contexts are underlined and
		the terms are in bold.
		For example,
		``[4]'' in article $a1$ denotes a citation to article $a4$
		with terms 
		$\tau_{1}$ and $\tau_{2}$.
	}
	\label{fig:fake_articles}
\end{center}
\end{figure}

\begin{figure}
\begin{center}
	\subfigure[
		Term labelled citation network: 
		$G(A, C)$
	]{
		\includegraphics[height=.40\linewidth]
			{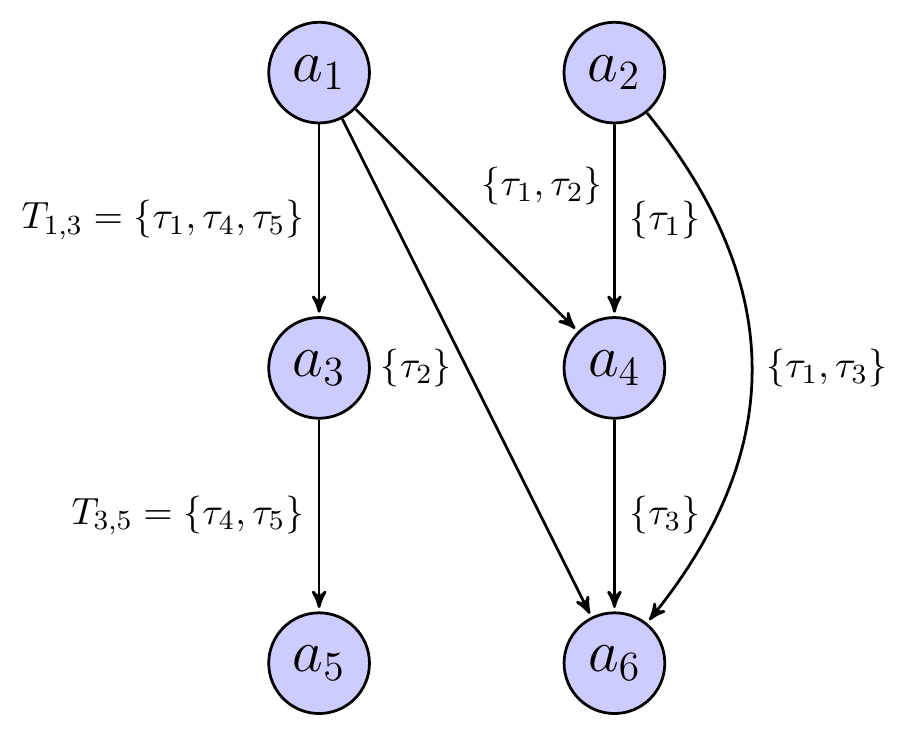}
		\label{fig:citationNetworkLabelled}
	} 
	\quad
	\quad
	\quad
	\subfigure[
		Citation network for $\tau_{1}$:
		$G_{\tau_{1}}(A, C_{\tau_{1}})$
	]{
		\includegraphics[height=.40\linewidth]
			{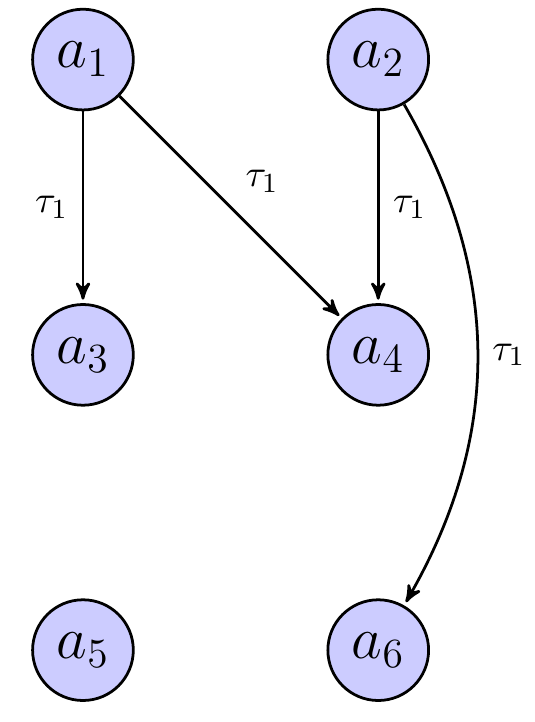}
		\label{fig:citationNetworkTermT1}
	} 
	\\ 
	\subfigure[
		Citation network for $\tau_{4}$:
		$G_{\tau_{4}}(A, C_{\tau_{4}})$
	]{
		\includegraphics[height=.40\linewidth]
			{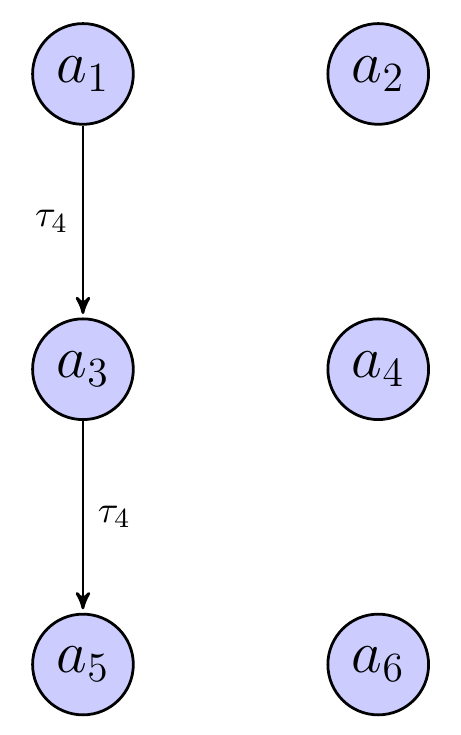}
		\label{fig:citationNetworkTermT4} 
	}
	\quad
	\quad
	\quad
	\subfigure[
		$\tau_{1}$-similar citation network:
		$G_{S_{\tau_{1}}}(A, C_{S_{\tau_{1}}})$
	]{
		\includegraphics[height=.40\linewidth]
			{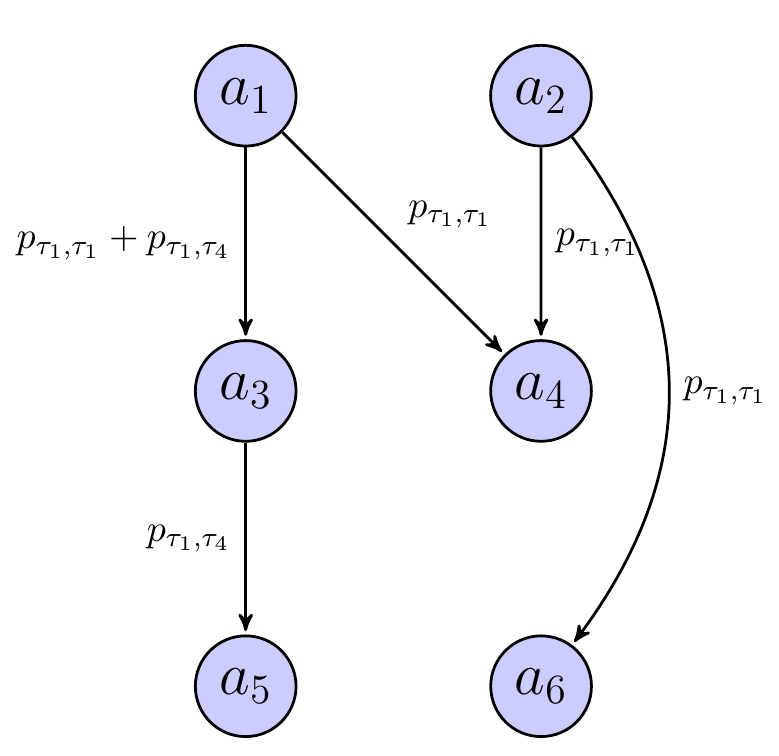}
		\label{fig:citationNetworkT1Similar}
	} 
	\\ 
	\subfigure[
		Article-term citation bipartite graph 
		(Weights of unlabeled edges are equal to $1$).
	]{
		\includegraphics[width=.85\linewidth]
			{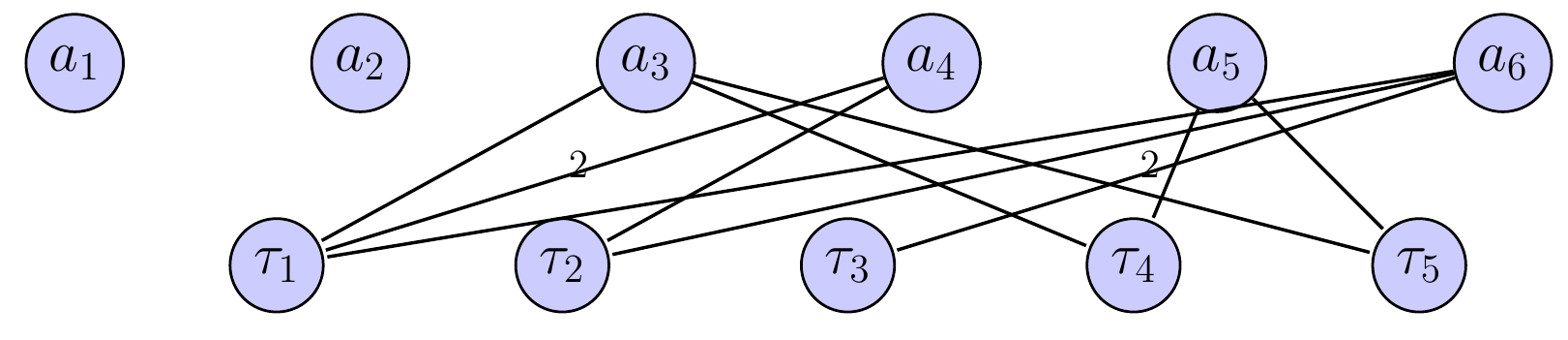}
		\label{fig:bipartiteNetwork}
	}
	\\ 
	\subfigure[
		Term Document Matrix $\hbMtrx{F}$.
	]{
		$\hbMtrx{F} = \bordermatrix{
			~ &a_{1} &a_{2} &a_{3} &a_{4} &a_{5} &a_{6} \cr
			\tau_{1} & 0 & 0 & 1 & 2 & 0 & 1 \cr 
			\tau_{2} & 0 & 0 & 0 & 1 & 0 & 1 \cr
			\tau_{3} & 0 & 0 & 0 & 0 & 0 & 2 \cr
			\tau_{4} & 0 & 0 & 1 & 0 & 1 & 0 \cr
			\tau_{5} & 0 & 0 & 1 & 0 & 1 & 0 \cr
		}$
		\label{fig:matrixF} 
	}
	\caption[Citation Networks.]{
		Citation Networks. 
		\subref{fig:citationNetworkLabelled} 
		Term labelled citation network $G(A, C)$. 
		\subref{fig:citationNetworkTermT1} 
		Citation network for $\tau_{1}$:
		$G_{\tau_{1}}(A, C_{\tau_{1}})$. 
		\subref{fig:citationNetworkTermT4} 
		Citation network for $\tau_{4}$:
		$G_{\tau_{4}}(A, C_{\tau_{4}})$. 
		\subref{fig:citationNetworkT1Similar} 
		$\tau_{1}$-similar citation network:
		$G_{S_{\tau_{1}}}(A, C_{S_{\tau_{1}}})$ 
		for the similar term set 
		$S_{\tau_{1}} = \left\{\tau_{1}, \tau_{4} \right\}$. 
		\subref{fig:bipartiteNetwork} 
		Article-term citation bipartite graph. 
		\subref{fig:matrixF} 
		Term document matrix $\hbMtrx{F}$.
	}
	\label{fig:all_in_one}
\end{center}
\end{figure}  

\emph{Citation context} is the text around the citation marker. 
The size of this text can be defined as a 
specific number of sentences, words or characters around the citation marker. 
We can form a \emph{citation network}, 
a directed graph, from citation information 
by creating directed edges from citing article to cited articles. 
Actually, 
an edge in a citation network carries more information than 
just a single binary relation. 
We can extract \emph{terms}, 
that is, a word or group of words,
from citation context which author used in order to explain the cited document.

Let $A$ be the set of all articles. 
We use lower case Latin letters for the articles in $A$ 
such as $i, j \in A$.
Let $T$ be the set of all terms used in all articles in $A$. 
In order to distinguish from articles,
we use lower case Greek letters for the terms in $T$ 
such as $\beta, \tau \in T$. 

A \emph{term-labelled citation network}, denoted by $G(A, C)$, 
is a directed graph with 
set of edges $C = A \times A$ where $(i, j) \in C$ 
if and only if
article $i$ cites article $j$. 
The edge $(i, j)$ is labelled with all terms in $T_{i j}$
where
$T_{i j} \subseteq T$ is the set of all terms 
that appear in 
at least one citation context in article $i$ to article $j$.
Note that we do not consider multiple occurrences of a term.
We have the term in $T_{ij}$ only once
although it is possible that 
the term can be repeated multiple times in the same citation context
or 
it might be present in multiple citations.
Note also that $T_{ij} = \emptyset$,
if there is no citation from article $i$ to article $j$,  
or
the citation context has no term in it.

Once we have term-labelled citation network,
we can obtain citation network specific to a term.
Let $\beta \in T$ be a term. 
The subgraph $G_{\beta}(A, C_{\beta})$ of $G(A, C)$ is called 
\emph{citation network for term-$\beta$} 
where 
$C_{\beta} \subseteq C$ and 
$(i, j) \in C_{\beta}$ 
if and only if 
$\beta \in T_{i j}$.

Note that the citation network $G(A, C)$ is the superposition of 
all term specific citation networks $G_{\beta}(A, C_{\beta})$ 
where $\beta \in T$.
That is,
$	
	C = \bigcup_{\beta \in T} C_{\beta}. 
$

As an example, 
suppose the entire repository of documents has six articles whose citation network is shown in \reffig{fig:fake_articles}. 
The corresponding term-labelled citation network with 
$A = \left\{ a_{1}, a_{2}, \cdots, a_{6} \right\}$ and 
$T =  \left\{ \tau_{1}, \tau_{2}, \cdots, \tau_{5} \right\}$
is given 
in \reffig{fig:citationNetworkLabelled}. 
The citation networks for terms $\tau_{1}$ and $\tau_{4}$ are
\reffig{fig:citationNetworkTermT1} and 
\reffig{fig:citationNetworkTermT4}, 
respectively.

Note in \reffig{fig:citationNetworkTermT4} 
that document $a_{5}$ is in the citation network for term $\tau_{4}$
although it does not contain $\tau_{4}$ in its text.
Note also that not only documents that are cited with term $\beta$,
but also the documents that cite them with $\beta$ 
are in the citation network for $\beta$ as in the case of $a_{1}$.
As long as $\tau_{4}$ is concerned, 
documents $a_{1}$, $a_{3}$ and $a_{5}$ are related while
documents $a_{2}$, $a_{4}$ and $a_{6}$ are not.
So one can disregard $a_{2}$, $a_{4}$ and $a_{6}$
since they are disconnected.

Once the citation network for term $\beta$ is obtained,
we can run standard network based ranking algorithms,
including 
in-degree,
HITS, and
PageRank,
and find important articles for this term.

As it is defined, 
we did not make any vertex reduction in the citation network for $\beta$.
Only arcs that are not $\beta$ related are removed.
As in the case of $a_{2}$ in \reffig{fig:citationNetworkTermT4},
removing arc leave some of the vertices \emph{isolated},
\hbIe
having both 0 in-degree and 0 out-degree.
We can remove the isolated vertices from the network 
while preserving all the information about $\beta$.
We call such networks as 
\emph{vertex reduced citation network for $\beta$}.
This reduction has important impact on performance of ranking algorithms.
See \refsec{sec:CitationNetworkForHadoop}.

\subsection{Term Similarity}

A term is not generally enough to describe fully 
a topic in scientific literature by itself 
and just using a single term is open to noises 
because of natural language usage such as synonyms etc. 
One of the key approaches of this work is that 
we use similar terms in the 
document retrieval process. 
This helps us to broaden the set of citation contexts 
we evaluate in the interested topic. 
Intuitively, 
two terms similar if 
they appear together in a considerable number of citations.
In order to formally define term similarity 
we need the following tools.

Term frequencies are related to articles 
by means of \emph{term document matrix},
denoted by $\hbMtrx{F} = [ f_{\beta j} ]$
where 
the entry $f_{\beta j}$ is the number of articles 
that cite article $j$
with term $\beta$ in their citation contexts.
That is, 
$f_{\beta j}$ is the indegree of article $j$ in the network $G_{\beta}$. 
Note that 
$\hbMtrx{F}$ is actually extracted from 
an undirected weighted bipartite graph 
between article nodes and term nodes.
As an example, 
\reffig{fig:bipartiteNetwork} 
is the bipartite network corresponding to
\reffig{fig:citationNetworkLabelled}
and 
\reffig{fig:matrixF} is the term document matrix.

We want to find similar terms, 
but especially the discriminative ones 
which are used to describe smaller set of articles. 
Simple term frequency has a problem that 
all terms are considered equally important, 
but certain terms have little or no discriminating power. 
For example, 
a collection of articles on ``cancer'' is likely to have 
the term ``cancer'' in nearly all citation contexts. 
So we decided to scale down the weights of terms 
which occur in lots of citation contexts. 
In principle, 
the idea is reducing term frequency weight of a term by a factor 
that grows with its citation context frequency it appears.
\emph{Term frequency-inverse document frequency} (\emph{tf-idf}) is a technique 
which is based on this 
idea~\cite{%
	Manning2008}. 
This method is widely used in information retrieval and text mining 
and it reflects how important a word is to a document in a collection. 
\emph{Inverse document frequency} 
for the term $\beta$ is defined by 
\[ 
	\idf(\beta) 
	= 
	\log \frac
			{|A|}
			{\sum_{j \in A} \sgn(f_{\beta j})}
\]
where $\sgn(x)$ is defined as
\[
	\sgn(x) 
	=
	\begin{cases}
		1, & x > 0, \\
		0, & x = 0, \\
		-1, & x < 0.
	\end{cases} 
\]
Obviously, $x < 0$ will not be the case in this context
since in-degree needs to be nonnegative.

We define \emph{weighted term document matrix} 
$\hbMtrx{N} = [n_{\beta j}]$ of size 
$|T| \times |A|$ by
\[
   \hbMtrx{N} = \hbMtrx{D} \hbMtrx{F}. 
\]
where
$\hbMtrx{D}$ $\mathit{= [d_{\tau \beta}]}$ is a 
$\mathit{|T| \times |T|}$ diagonal matrix defined by
\[
	d_{\beta \tau} 
	=
	\begin{cases}
		\idf(\beta), &\tau = \beta, \\
		0, &\text{otherwise}. 
	\end{cases} 
\]

We are one step away to define similarity of terms.
Let 
$\vecBeta$ 
and 
$\vecTau$ 
be the row vectors 
corresponding to terms
$\beta$
and 
$\tau$ 
in $\hbMtrx{N}$,
respectively. 
Entries of 
$\vecBeta$ 
and 
$\vecTau$ 
show the respective weighted term frequencies of terms 
$\beta$
and 
$\tau$ 
for the articles in the dataset. 
If somebody wants to find out how much article coverages of these terms overlap, 
then he needs to compare corresponding row vectors 
$\vecBeta$ 
and 
$\vecTau$. 

For this purpose we use sample Pearson correlation $p_{\beta \tau}$ of 
$\vecBeta$ 
and 
$\vecTau$. 
Let 
$
\hbVect{x} = [x_{1}, \cdots, x_{n}], 
\hbVect{y} = [y_{1}, \cdots, y_{n}] 
\in \mathbb{R}^{n}
$ 
be vectors with $n$ entries.
Then \emph{sample Pearson correlation coefficient} of 
$\hbVect{x}$ and 
$\hbVect{y}$ is given as 
\[
	p_{x y} 
	= \frac
	{
			\sum_{i=1}^{n} 
			(x_{i} - \hbAvrg{\hbVect{x}}) 
			(y_{i} - \hbAvrg{\hbVect{y}}) 
	}
	{
		\sqrt{
			\sum_{i=1}^{n} 
			(x_{i} - \hbAvrg{\hbVect{x}})^{2} 
			} 
		\sqrt{
			\sum_{i=1}^{n} 
			(y_{i} - \hbAvrg{\hbVect{y}})^{2}
			}
	}
\]
where 
$\hbAvrg{\hbVect{x}}$
is the average of entries of vector $\hbVect{x}$.
The sample Pearson correlation coefficient is 
a measure of the linear correlation 
between two samples $\hbVect{x}$ and $\hbVect{y}$, 
giving a value between $-1$ and $1$ inclusive. 
A value of $1$ means that a linear equation describes 
the relationship between $\hbVect{x}$ and $\hbVect{y}$, 
with all data points lying on a line where $\hbVect{y}$ increases as $\hbVect{x}$ increases. 
A value of $-1$ means that all data points lie on a line 
for which $\hbVect{y}$ decreases as $\hbVect{x}$ increases. 
This case is irrelevant for our dataset, because to get a value of $-1$ 
for two terms $\tau$ and $\beta$, 
they have to be complement of each other. 
This is not probable on a large collection of articles. 
A value of $0$ means that there is no linear correlation between the samples. 

Finally we can define term similarity.
Term $\tau$ is said to be \emph{similar} to term $\beta$ 
if and only if
$p_{\beta \tau} < \delta$
for some
$0 < \delta < 1$. 
Note that 
$\delta$ is a cross validation parameter and 
its value changes among topics. 
In our experiments, we found the optimum value by trial and error.

\subsection{Term Similar Citation Network}
\label{sec:TermSimilarCitationNetwork}

For a given term $\beta$, 
we define \emph{$\beta$-similar term set} $S_{\beta}$
as
\[
	S_{\beta} 
	= 
	\left\{ \tau \in T \; | \; \tau \text{ is similar to } \beta \right\}.
\]
Note that 
$S_{\beta}$ is nonempty since $\beta \in S_{\beta}$.

Now, 
we can combine citation networks for similar terms 
into one directed, labelled network.
The subgraph $G_{S_{\beta}}(A, C_{S_{\beta}})$ of $G(A, C)$ is called 
\emph{$\beta$-similar citation network} 
where
we combine edges in the citation networks of the terms 
that are similar to $\beta$, 
i.e.,
$C_{S_{\beta}} = \bigcup_{\tau \in S_{\beta}} C_{\tau}$.
Weight $w_{i j}$ for the edge $(i, j) \in C_{S_{\beta}}$ 
is the sum of weights of the edges combined,
i.e., 
\[
	w_{i j}
	=
	\sum_{(i, j) \in T_{i j} \bigcap S_{\beta}}  p_{\beta \tau}.
\]

For example consider \reffig{fig:citationNetworkLabelled}. 
Assume that we calculated similarity set 
$S_{\tau_{1}} = \left\{ \tau_{1}, \tau_{4} \right\}$ 
for the term 
$\tau_{1}$.
Then 
$\tau_{1}$-similar citation network,
given in \reffig{fig:citationNetworkT1Similar},
would be superposition of networks 
$G_{\tau_{1}}(A, C_{\tau_{1}})$
and
$G_{\tau_{4}}(A, C_{\tau_{4}})$ 
given in 
\reffig{fig:citationNetworkTermT1}
and
\reffig{fig:citationNetworkTermT4},
respectively.

After forming $\beta$-similar citation network for a given term $\beta$, 
we can run common ranking algorithms on this network and 
find the most important articles for the topic 
represented by the term $\beta$. 

A note about performance has to be made here.
Rather than citation networks for terms,
if we use vertex reduced citation networks for terms,
then the resulted $\beta$-similar citation network 
would be much smaller in number of vertices, 
yet it has all the information necessary for network based ranking.
This is an important performance benefit for real life cases.
See \refsec{sec:CitationNetworkForPowerLaw}.

\section{Evaluation of Results}

We implement the approach given in \refsec{sec:Methodology} on a CiteSeerX dataset.
Obtain some sample runs.

\subsection{CiteSeerX Dataset}

``SeerSuite is a framework for scientific 
and academic digital libraries and search engines built by crawling scientific 
and academic documents from the web 
with a focus on providing reliable, robust services. 
In addition to full text indexing, 
SeerSuite supports autonomous citation indexing 
and automatically links references in research articles 
to facilitate navigation, analysis and 
evaluation''~\cite{%
	Teregowda2010}. 
CiteSeerX, 
where we get our dataset, 
is an instance of SeerSuite. 
CiteSeerX team was nice enough to provide us a snapshot of June 2012.
The data contains nearly 1.8 million scientific articles and 
41.5 million citation contexts extracted from these articles.

In our dataset, 
citation contexts are marked with their citations. 
On the other hand, 
we still need to identify which terms in a citing paper refer to 
which of its citations. 
This is not an easy problem due to nature of different citation techniques. 
Ritchie~\hbEtal stated some of the problems 
with matching terms in citation context 
with correct 
citations~\cite{%
	Ritchie2006}: 
\begin{itemize}
	   
	\item 
	Length of text, 
	which refers to citations, 
	differentiate from article to article.
	   
	\item 
	If citation context blocks are physically very close to each other, 
	then citation terms for different citations would be overlapped.
	   
	\item 
	Citation marks may appear in different places 
	such as some citation texts start with citation mark 
	while others end with citation mark.
	   
	\item 
	Some citation contexts may contain contradictory terms for cited articles.
\end{itemize}

\subsection{Term Identification}
\label{sec:TermIdentification}

As explained in \refsec{sec:Methodology},
our algorithm works on terms 
which could be composed of one, two or more words. 
We use CiteSeerX specific tools to come up with 
the following simple scheme for term identification 
in citation context~\cite{%
	Teregowda2010,%
	Councill2008}.
But any other, 
possibly more sophisticated technique, 
would only do better.
 
The words that are used to describe the cited paper 
will stand close to the citation marker, 
so we used a window of fixed size for citation context like previous 
studies~\cite{%
	Bradshaw2003}. 
Citation contexts in our dataset consist of around 400 characters 
which are equally divided to both sides of citation marker
for which we use ParsCit open 
library~\cite{%
	Councill2008}.

The CiteSeerX database~\cite{%
	Teregowda2010}
provides a keyword list for each paper.
We will use a keyword,
which could be single word or multiple words, 
as a term.
We collect all the terms of all papers into a \emph{set of terms} $T$.
Any word sequence in citation context is check against this set.
If a sequence is in the set, 
we simply take it as a \emph{term}.

It is possible that a concept and its inventor go 
together in citation context. 
So there is a potential problem of how to handle such cases.
As an example, 
consider Game of Life invented by 
Conway~\cite{%
	Gardner1970}. 
One expects that a document on Game of Life 
would have a keyword ``game of life'', 
and hopefully no keyword ``conway''. 
So our approach would not consider a combination of
``game of life'' and ``conway''	as a term.
If inventor and the concept together become a keyword,
than our system would take that as a term, 
as it correctly should.

Note that in this definition, 
a term can be a single word, bigram, or composed of three or more words.
We consider only single words and bigrams for our investigation.
Once we identify a term, 
which could be a single word, or composed of multiple words,
our approach would be the same.
Therefore term identification in general, 
or bigram selection in particular, 
would not change the contribution of our approach.

\subsection{Citation Network for the Term ``hadoop''}
\label{sec:CitationNetworkForHadoop}

\begin{table*}
\begin{center}
	\caption{Ranking based on Citation Network for ``hadoop''.}
	{\footnotesize 	
		\begin{tabular}{|l|r|r|r|}
			\hline
			\textbf{Paper \textbackslash Ranking Method}
			&\textbf{In} 
			&\textbf{HITS}
			&\textbf{Page}\\ 
			%
			&\textbf{Degree} 
			&
			&\textbf{Rank}\\ 
			\hline
			MapReduce: Simplified Data Processing on $\cdots$
			\cite{Dean2004}
			&1 &1 &1 \\ 
			The Google File System
			\cite{Ghemawat2003}
			&2 &2 &2 \\ 
			Evaluating MapReduce for Multi-Core and $\cdots$ 
			\cite{Ranger2007}  
			&3 &3 &3 \\ 
			Pig Latin: a Not-So-Foreign Language for  $\cdots$
			\cite{Olston2008}
			&4 &4 &4 \\ 
			\hline
		\end{tabular}
	}
	\label{tbl:rankingHadoop}
\end{center}
\end{table*} 

\begin{figure}[htbp] 
\begin{center}
	\fbox{
		\includegraphics[width=\hbFigSize\columnwidth]%
			{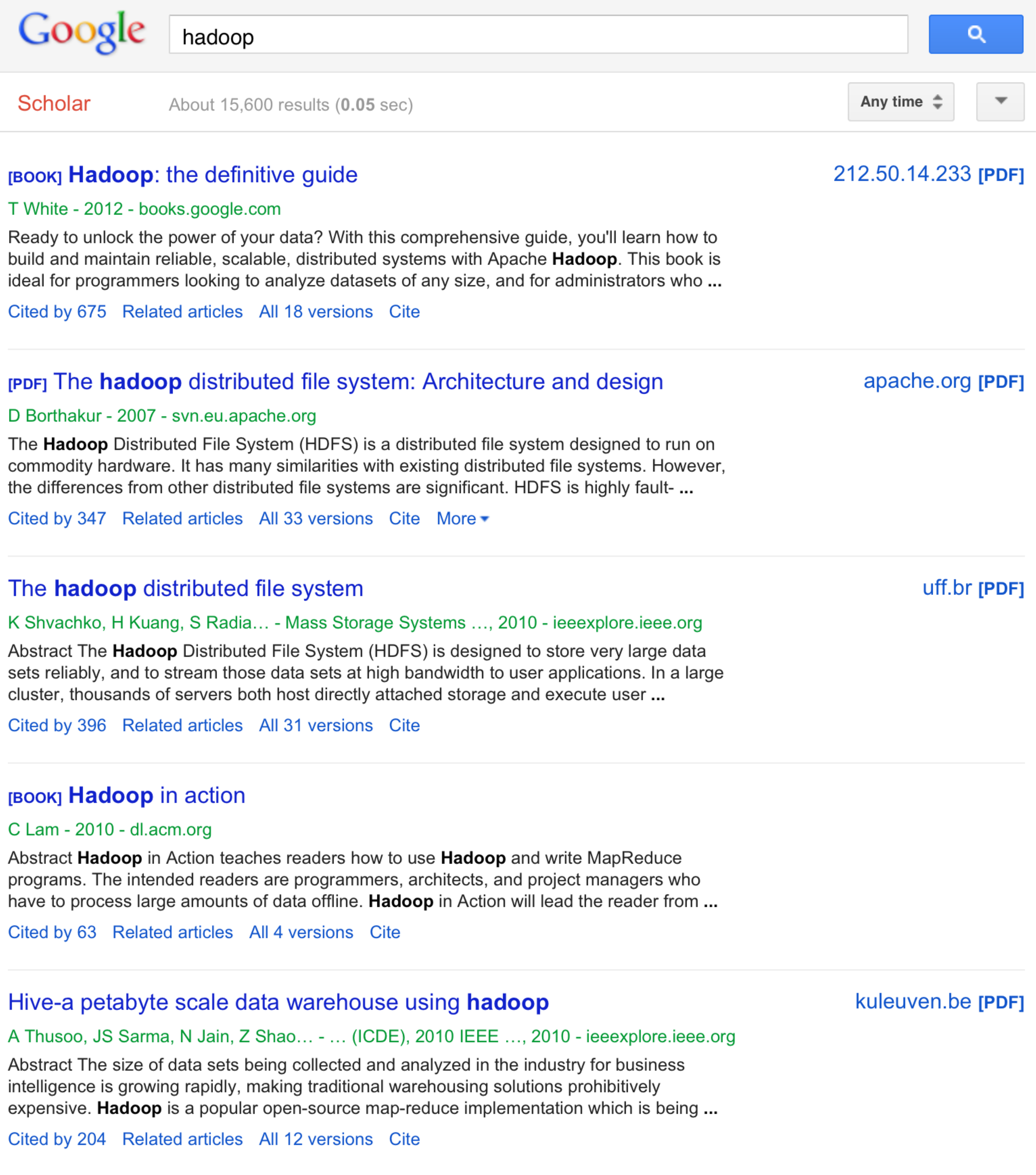}
	}
	\caption{Results for ``hadoop'' on Google Scholar.}
	\label{fig:hadoopGoogleScholar}
\end{center}
\end{figure}

\begin{figure}[htbp] 
\begin{center}
	\fbox{
		\includegraphics[width=\hbFigSize\columnwidth]%
			{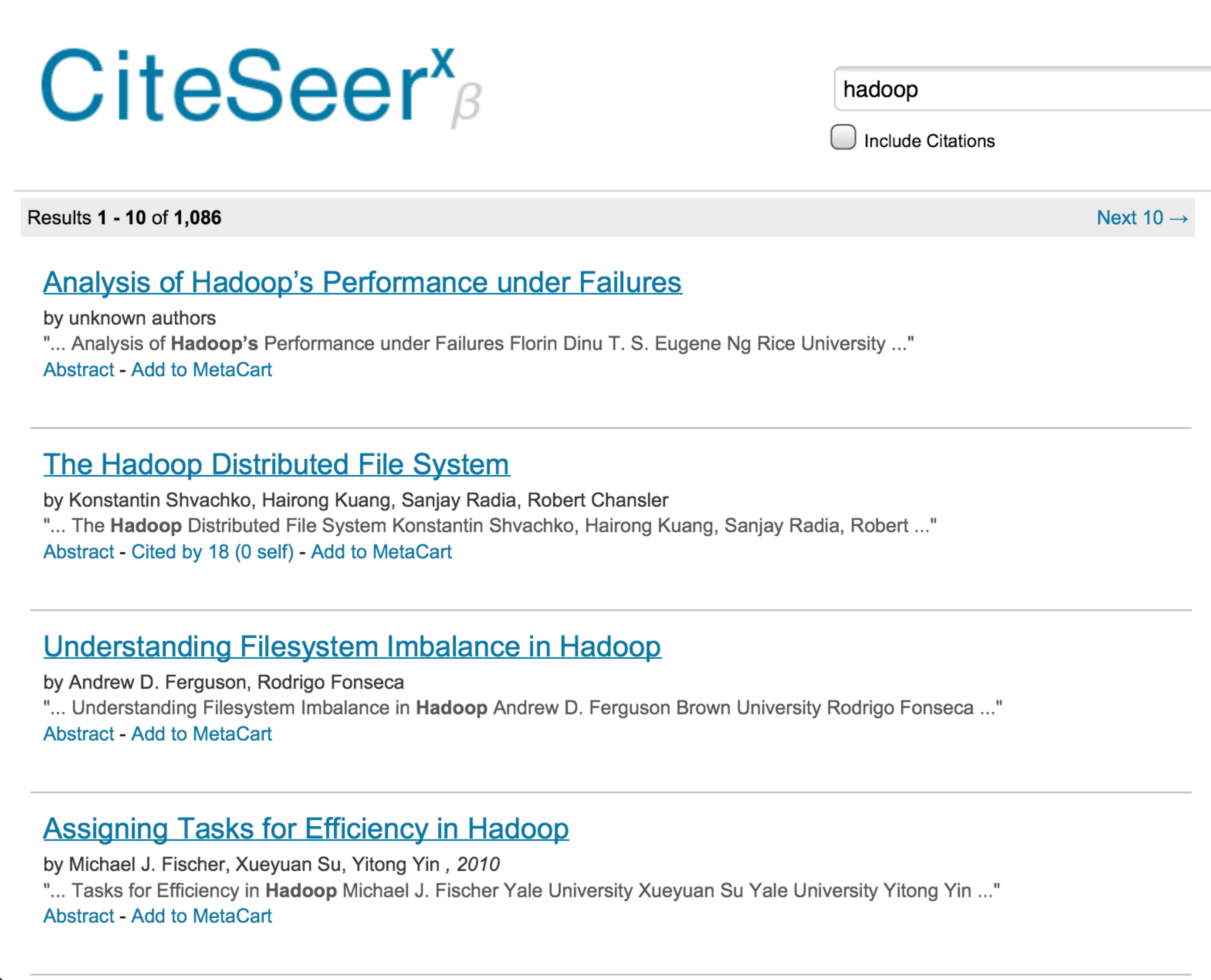}
	}
	\caption{Results for ``hadoop'' on CiteSeerX.}
	\label{fig:hadoopCiteseerx}
\end{center}
\end{figure}

Our approach not only finds related documents 
but also ranks them properly
even if the term is not in the document.
We investigate this in the case of ``Hadoop''.
Hadoop was derived from Google File System (GFS)~\cite{%
	Ghemawat2003} 
and Google's MapReduce~\cite{%
	Dean2004} 
papers. 
These papers published respectively in 2003 and 2004 
while Hadoop term is coined in 2005. 
Our method detected these articles as the highest ranking papers as 
it is seen in \reftbl{tbl:rankingHadoop},
while existing search engines 
Google Scholar and CiteSeerX 
do not even list them.
The search results for ``hadoop'' are given in
\reffig{fig:hadoopGoogleScholar} 
and
\reffig{fig:hadoopCiteseerx}  
for Google Scholar and CiteSeerX, respectively.

The descriptive case given in \reffig{fig:all_in_one} will be useful.
Consider term $\tau_{1}$ in \reffig{fig:citationNetworkTermT1}.
Note that document $a_{4}$ is in the citation network for term $\tau_{1}$
although it does not contain $\tau_{1}$.
Documents 
$a_{1}$,
$a_{2}$,
$a_{3}$,
$a_{4}$,
$a_{6}$
are related to $\tau_{1}$ 
but 
$a_{5}$ is not.
Compared to the entire network given in \reffig{fig:citationNetworkLabelled},
the network in \reffig{fig:citationNetworkTermT1}
is reduced in size 
since there is no point including document $a_{5}$ to the network.

``Hadoop'' itself is a very descriptive term, 
so we didn't include similar terms.
We consider all the citation context of 1.8 million articles in our data set.
We pick the ones that contain ``Hadoop'' as a term 
in their citation contexts.
Similar to the citation network for $\tau_{1}$ 
given in \reffig{fig:citationNetworkTermT1},
we obtain the citation network for ``Hadoop'', 
which has 752 articles only.
Then, 
we apply the usual in-degree, HITS and PageRank algorithms for ranking on 
the citation network for ``Hadoop''.
Hence \reftbl{tbl:rankingHadoop} is obtained.
One should note that vertex reduction has a great impact of 
reducing the number of vertices from 1.8 million to 752.

\subsection{Citation Network for ``power law''}
\label{sec:CitationNetworkForPowerLaw}

Following the outline given in \reffig{fig:all_in_one},
we investigate ``power law''-similar citation network,
which corresponds to \reffig{fig:citationNetworkT1Similar}.
Similar to \reffig{fig:citationNetworkTermT1},
we start with the citation network for ``power law'',
which contains $12,616$ nodes.
We will enrich this network with the networks of ``power law''-similar terms.
First we get ``power law''-similar terms.
Terms with ``power law''-similarity more than a threshold of $0.35$ 
are listed in 
\reftbl{tbl:similarTermsPowerLaw}.
Then we obtain citation network for each term 
in \reftbl{tbl:similarTermsPowerLaw}.
By using superposition of these term networks 
with the citation network for ``power law'', 
we create 
the ``power law''-similar citation network,
which consists of $16,487$ nodes.
Running in-degree, HITS, Pagerank or any other ranking algorithm on a network of this sizes computationally is not a problem.

Here again, 
we should emphasis the importance of vertex reduction to obtain networks with 
$12,616$ and
$16,487$ nodes.
Note also that ``power law''-similarity
enrichment adds $3,871$ nodes to the citation network for ``power law''.

\begin{table}[thbp] 
\begin{center}
	\caption{The set of ``power law''-similar terms, $S_{\text{power-law}}$.}
	\begin{tabular}{|l|r|}\hline 
		\textbf{Term}& \textbf{Similarity Score}\\ \hline
		power law & 1.00 \\ 
		degree distribution & 0.83 \\ 
		web graph & 0.56 \\ 
		preferential attachment & 0.45 \\ 
		scale free & 0.38 \\ 
		\hline
	\end{tabular}
	\label{tbl:similarTermsPowerLaw}
\end{center}
\end{table} 

\begin{figure}[htbp] 
\begin{center}
	\fbox{
		\includegraphics[width=\hbFigSize\columnwidth]%
			{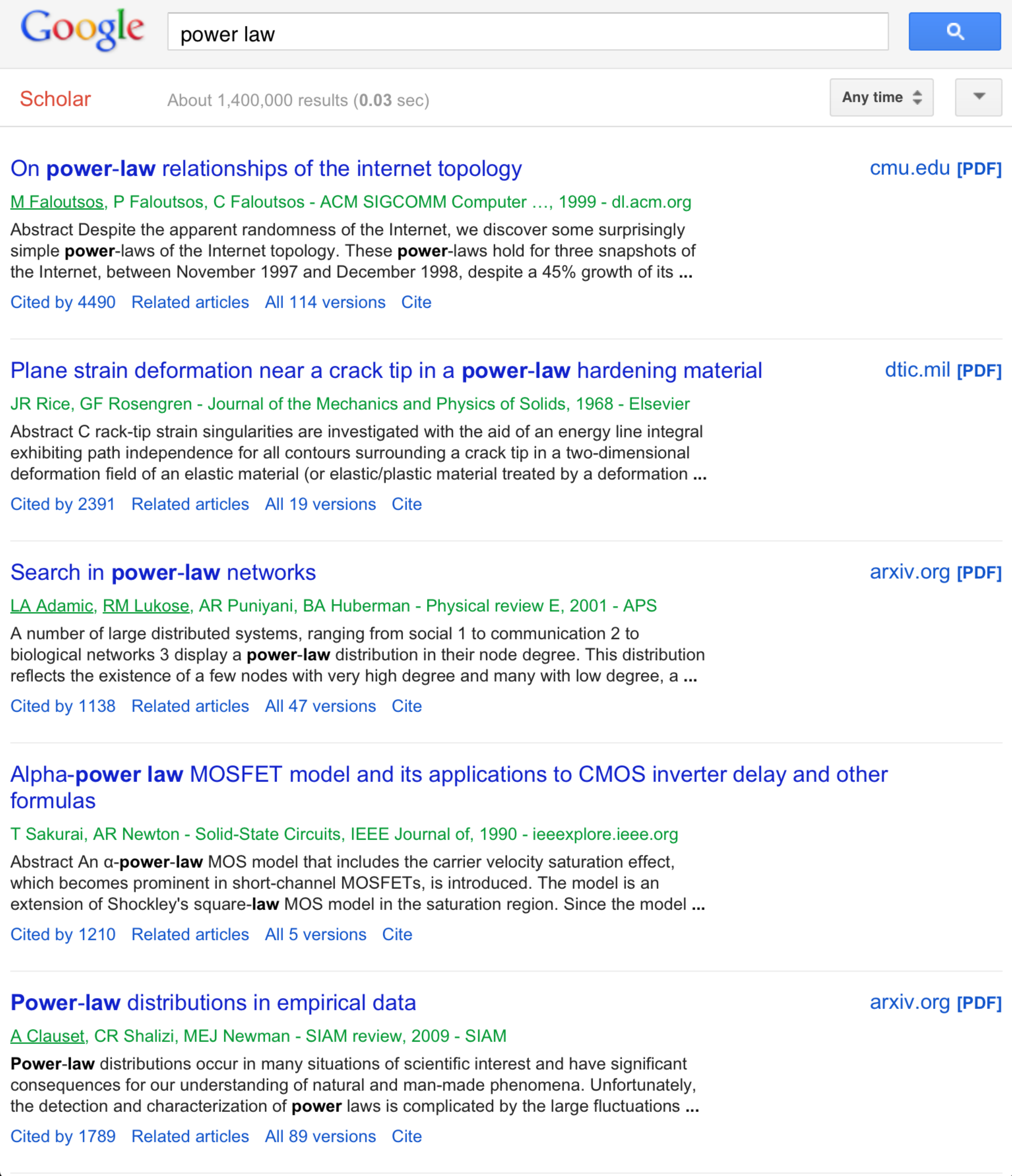}
	}
	\caption{Results for ``power law'' on Google Scholar.}
	\label{fig:power_law_google_scholar_figure}
\end{center}
\end{figure}

\begin{figure}[htbp] 
\begin{center}
	\fbox{
		\includegraphics[width=\hbFigSize\columnwidth]%
			{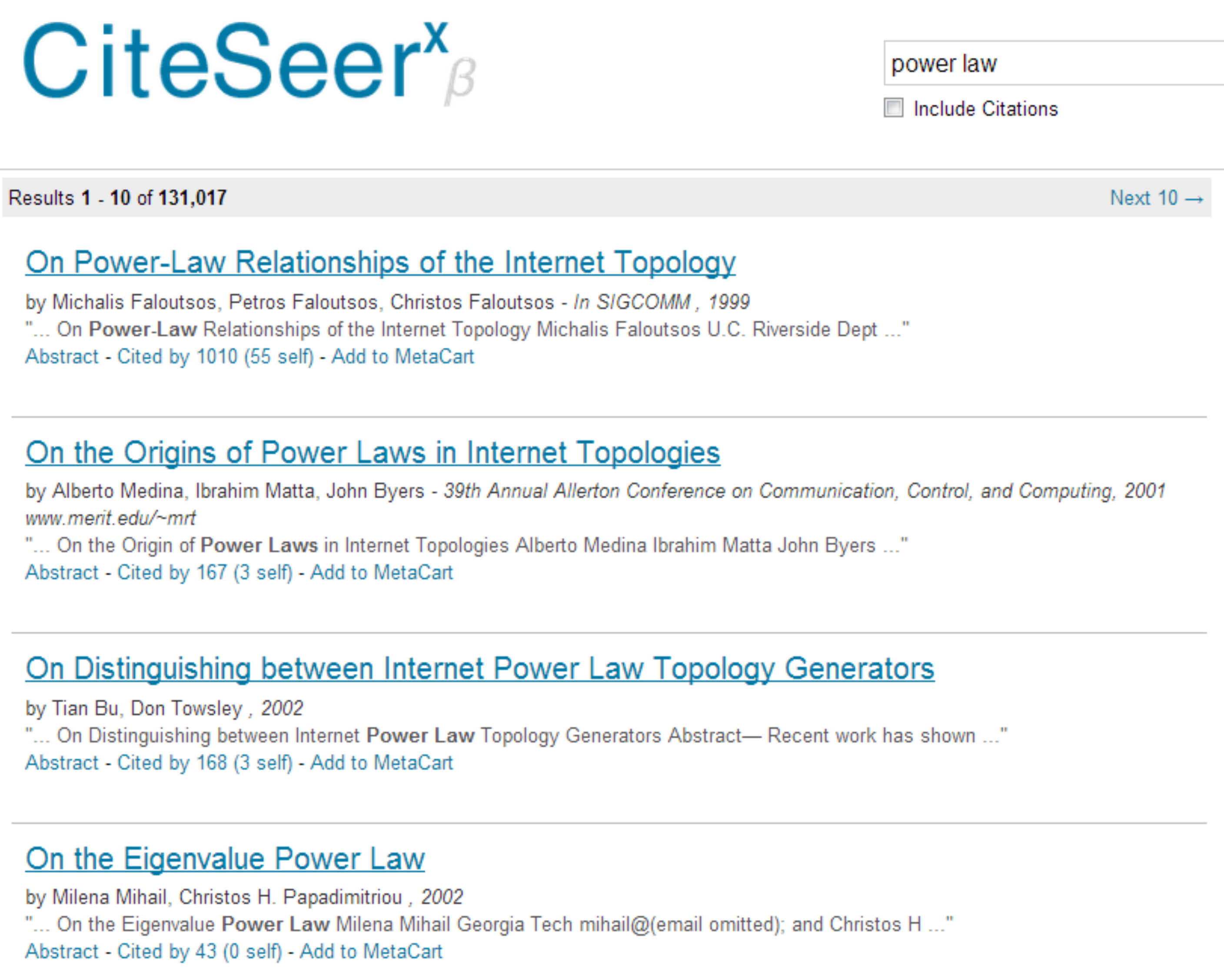}
	}
	\caption{Results for ``power law'' on CiteSeerX.}
	\label{fig:power_law_citeseerx_figure}
\end{center}
\end{figure}

Ranking results are reported at 
\reftbl{tbl:rankingPowerLaw} 
for the articles which take place in the top ten for all methodologies
namely, 
in-degree, 
HITS~\cite{%
	Kleinberg1999} 
and 
PageRank~\cite{%
	Brin1998}. 
As it is shown on the table, 
we are able to identify most prominent articles 
in ``power law'' related topic.
We also searched ``power law'' term on academic literature search engines and 
the top results are shown in 
\reffig{fig:power_law_google_scholar_figure} 
for Google Scholar~\cite{urlGoogleScholar}
and in 
\reffig{fig:power_law_citeseerx_figure} 
for CiteSeerX. 
We can comfortably say that our results consist of more related 
and important articles for the topic represented by ``power law''.

\begin{table*}[thbp] 
\begin{center}
	\caption{
		Ranking of articles based on ``power law''-Similar Citation Network.
	}
	{\footnotesize 
		\begin{tabular}{|l|r|r|r|} 
			\hline
			\textbf{Paper \textbackslash \ Ranking Method}
			&\textbf{In} 
			&\textbf{HITS}
			&\textbf{Page}\\ 
			%
			&\textbf{Degree} 
			&
			&\textbf{Rank}\\ 
			\hline
			Emergence of Scaling in Random Networks 
			\cite{Barabasi1999} 
			&1&1&1 \\ 
			On Power-Law Relationships of the Internet Topology   
			\cite{Faloutsos1999} 
			&2&2&2 \\ 
			Statistical Mechanics of Complex Networks 
			\cite{Albert2002} 
			&3&3&6 \\ 
			Collective Dynamics of `small-world' Networks 
			\cite{Watts1998} 
			&5&4&9 \\ 
			The Structure and Function of Complex Networks 
			\cite{Newman2003} 
			&6&5&10 \\ 
			Alpha-Power Law MOSFET Model 
			and its $\cdots$   
			\cite{Sakurai1990} 
			& 4 & 6 & 10 \\ 
			\hline		  
		\end{tabular}
	}
	\label{tbl:rankingPowerLaw}
\end{center}
\end{table*} 

We observe the most obvious benefit of using citation context with the appearance of article 
``Collective Dynamics of `small-world' Networks''~\cite{%
	Watts1998} 
in our list of \reftbl{tbl:rankingPowerLaw}. 
The article introduces the concept of ``small world'',
and
\reftbl{tbl:similarTermsPowerLaw} does not contain ``small world''.
It contains 
neither the term ``power law'' 
nor any other terms in the set of 
``power law''-similar terms 
of \reftbl{tbl:similarTermsPowerLaw}. 
However, this is an important article for this topic 
and Google Scholar and CiteSeerX miss it while we find it.

\begin{table}[thbp] 
\begin{center}
	\caption{
		The ranking of the most descriptive terms of the top four articles 
		in the ``power law''-Similar Citation Network.
	}
	\begin{tabular}{|l|r|r|r|r|}
		\hline
		\textbf{Term \textbackslash Article}
			& \textbf{\refcite{Barabasi1999}} 
			& \textbf{\refcite{Faloutsos1999}}
			& \textbf{\refcite{Albert2002}} 
			& \textbf{\refcite{Watts1998}}\\ 
		\hline
		power law                &1 &1 &2 &6 \\
		scale free               &2 &4 &1 &5 \\ 
		preferential attachment  &3 &16&8 &9 \\ 
		degree distribution		 &4 &2 &3 &7 \\ 
		random graphs            &5 &5 &6 &4 \\ 
		small world              &6 &7 &5 &1 \\ 
		social networks          &7 &6 &7 &3 \\ 
		complex networks         &8 &10&4 &8 \\ 
		web graph				 &9 &8 &26&34\\ 
		internet topology        &10&3 &14&26\\ 
		clustering coefficient   &16&25&8 &2 \\ 
		\hline  
	\end{tabular}
	\label{tbl:descriptiveTerms}
\end{center}
\end{table} 

We further investigate the relation of terms ``small world'' 
and ``power law'' as follows.
For an article,
we can use its citation contexts to calculate tf-idf values of the terms 
that describe it.
Since a term with higher tf-idf value describes the article better,
the terms are ordered in tf-idf values. 

The ranking of the most descriptive terms for the top four articles in 
``power law''-similar citation network is given 
in \reftbl{tbl:descriptiveTerms}.
Since \refcite{Barabasi1999} introduces the term ``scale free'',
the most descriptive terms of it are 
``power law'',
``scale free'',
``preferential attachment'', and
``degree distribution'',
as expected. 
But the sixth most descriptive term is ``small world''.

Concept of ``small world'' is introduced by \refcite{Watts1998} in 1998.
As expected terms 
``small world'' and 
``clustering coefficient''
are the top terms for \refcite{Watts1998}.
Interestingly, 
``scale free'' is the number five term for \refcite{Watts1998}
although the concept of
``scale free'' networks is introduced one year later 
by \refcite{Barabasi1999}.

\section{Conclusion}

Citation indexes are generally based on Boolean retrieval, 
so every article using a set of query terms 
is equally likely to be listed for the given query. 
The author of an article uses many words while explaining her research 
that may contain words not related with main contributions of the article. 
Hence, unrelated articles may rank highly in search results for a query, 
simply because they are important articles in another area 
and contain the query terms. 
So there is a need for a system which is able to measure both relevance 
and impact.

We are interested in finding fundamental 
and important documents in a context sensitive way. 
We especially target scientific literature 
because of the existing potential of citation structure. 
The text around citation marks represents very concise information 
about cited documents. 
Probably, the most prominent criticism is that 
citation analysis based on raw citation counts ignores 
the underlying reasons for the citation. 
So, we come up with a solution for this problem.

In this work, 
we presented a method to utilize citation contexts 
in order to rank important articles in a topic specific way. 
For a given  term which represents the interested topic, 
first we formed a set of similar terms. 
Then we detected citation contexts which contain terms from this set. 
Only by using detected citations we created topic specific networks. 
Finally, we applied common link analysis methods in order to find 
the most prominent articles in these topic specific citation networks.

It would draw attention of someone that 
we did not report the ranking for hub scores 
while we reported authority scores for the HITS algorithm. 
This is because we did not find meaningful results for hub scores 
where most of the nodes had a score of zero. 
According to our observations in these experiments, 
the reason for this is that 
there are no articles 
which list most of the prominent articles more than others.

This is an unsupervised problem and evaluating results 
require knowledge in target context. 
So we kept our test cases limited to domain of complex networks,
in which we can interpret our findings.
As a future work, 
we can work with academicians from different research topics 
in order to evaluate our system broadly.

Optimization of the algorithms for performance is 
another issue for the future.
We have not investigated the scalability of our approach 
for very large data sets such as www network.

There are limitations of our approach.	
First of all, it needs field test.
That is, 
it has to be 
used by real people,
in a real life environment, 
on real life data
which we are unfortunately unable to do.
Then quality of the results can be judge better.
Also there would be computational issues to be solved 
such as 
handling very large data, or
handling massive queries such as Google Scholar experiences.
Note that our method relies on external term identification.
CiteSeerX dataset provides us terms as keywords.
We would need term identification method
if we were to apply our method to some other datasets, 
such as legal documents.

\section*{Acknowledgments}
	The authors would like to thank 
	CiteSeerX team for providing the data as
	a snapshot of June 2012.
	We also thank to our anonymous reviewers 
	for pointing out unclear points in our draft manuscript.
	This work was partially supported by the 
	Turkish State Planning Organization (DPT) TAM Project (2007K120610).


\bibliographystyle{IEEEtran}
\bibliography{ContextSearch}

\end{document}